\documentclass[english]{article}
\usepackage[T1]{fontenc}
\usepackage[latin9]{inputenc}
\usepackage{amsmath}
\usepackage{amssymb}

\makeatletter
\newcommand{\lyxaddress}[1]{
\par {\raggedright #1
\vspace{1.4em}
\noindent\par}
}

\makeatother

\usepackage{babel}
\begin{document}

\title{\textbf{Ideal gas with a varying (negative absolute) temperature:
An alternative to dark energy?}}

\author{\textbf{Subhajit Saha$^{1},$ Anindita Mondal$^{2}$ and Christian
Corda$^{3}$}}
\maketitle

\lyxaddress{$^{1}$ Department of Mathematics, Panihati Mahavidyalaya, Kolkata
700110, West Bengal, India, e-mail: $subhajit1729@gmail.com$.}

\lyxaddress{\textbf{$^{2}$}Department of Astrophysics and Cosmology, S. N. Bose
National Centre for Basic Sciences, JD Block, Sector III, Salt Lake
City, Kolkata 700106, India, e-mail: $aanyndeta@gmail.com$.}

\lyxaddress{$^{3}$ Research Institute for Astronomy and Astrophysics of Maragha
(RIAAM), P.O. Box 55134-441, Maragha, Iran, e-mail: $cordac.galilei@gmail.com$. }
\begin{abstract}
The present work is an attempt to investigate whether the evolutionary
history of the Universe from the offset of inflation can be described
by assuming the cosmic fluid to be an ideal gas with a specific gas
constant but a varying negative absolute temperature (NAT). The motivation
of this work is to search for an alternative to the \textquotedbl{}exotic\textquotedbl{}
and \textquotedbl{}supernatural\textquotedbl{} dark energy (DE). In
fact, the NAT works as an ``effective quintessence'' and there is
need to deal neither with exotic matter like DE nor with modified
gravity theories. For the sake of completeness, we release some clarifications
on NATs in Section 3 of the paper.
\end{abstract}
\begin{quote}
\textbf{Keywords: Negative absolute temperature; Cosmic acceleration;
Ideal gas law; Dark energy}

\textbf{PACS Numbers: 98.80.-k}
\end{quote}

\section{Introduction}

Since the intriguing discovery of the late time cosmic acceleration
in 1998, which was expected by the Indian physicist B. Sidharth \cite{key-1}
before the cosmological observations of the High-­Z Supernova Search
Team \cite{key-2} and of the Supernova Cosmology Project \cite{key-3},
there have been rigorous attempts to incorporate this unexpected observations
into standard cosmology. This is the famous DE issue, which is a challenging
problem in cosmology. The simplest and the widely accepted form of
DE is the so-called cosmological constant, which arises from a historical
idea of Einstein in a different context \cite{key-4}. However, it
is plagued by the coincidence \cite{key-5} and the cosmological constant
\cite{key-6} problems. As a consequence, alternative ways have been
proposed such as modified gravity theories {[}7 - 9{]}, inhomogeneous
cosmological models \cite{key-10,key-11} etc. But, again, each one
of them comes with their own disadvantages. Other models such those
of particle creation have also been investigated in the past \cite{key-12}
as well as recent times \cite{key-13}. The advantage of these models
is that, in order to explain the evolutionary history of the cosmos,
one needs neither any exotic matter nor any modification of Einstein's
general theory of relativity (GTR). Moreover, such models seem to
be thermodynamically motivated. Nevertheless, the exact form of the
particle creation rate have still not been identified. Thus, the research
in this particular field has been largely phenomenological. 

For the sake of completeness we recall that there are also some new
attempts introducing new origins for the DE problem. The probable
non-extensive features of spacetime {[}27\textendash 29{]}, and the
tendency of spacetime to couple with matter in a non-minimal way \cite{key-30}
are some of these attempts. Moreover, the negative temperature of
cosmic fluid can also be obtained as the direct result of solving
the Friedmann and Thermodynamics equations simultaneously \cite{key-31}. 

Another key point of the current cosmological tapestry is the inflationary
era, i.e. the idea that after the initial singularity the Universe
was leaded by a very fast phase of expansion, which was due to a big
negative pressure \cite{key-14,key-15}. 

In this paper we attempt to investigate whether the evolutionary history
of the Universe from the offset of inflation can be described by assuming
the cosmic fluid to be an ideal gas with a specific gas constant but
a varying NAT. In this regard, on one hand it is worthwhile to mention
that NAT has been considered in the context of Cosmology before \cite{key-16}.
On the other hand, we also stress that the assumption that the cosmic
fluid should be an ideal gas at the end of the inflationary era has
been used in \cite{key-17} while in \cite{key-18} it has been shown
that an ideal gas cosmological solution leads to a universal accelerated
expansion which seems consistent with the supernova observations.
Thus, the most important motivation behind this work is that negative
absolute temperature gives rise to negative pressures which can therefore
behave as DE and lead the Universe to accelerate. 

\section{The ideal gas model with negative absolute temperature}

Let us consider a flat Friedmann-Lemaitre-Robertson-Walker (FLRW)
Universe with the line element (we work with $c=1$ in the following)
\cite{key-19} 
\begin{equation}
ds^{2}=-dt^{2}+a^{2}(t)\left[dr^{2}+r^{2}(d\theta^{2}+\text{sin}^{2}\theta d\phi^{2})\right].
\end{equation}
The FLRW field equations are
\begin{equation}
H^{2}=\frac{8\pi G}{3}\rho~~~~~~~~\text{and}~~~~~~~~\dot{H}=-4\pi G(\rho+p),\label{efe}
\end{equation}
from which one can derive the conservation equation as 
\begin{equation}
\dot{\rho}+3H(\rho+p)=0.\label{ce}
\end{equation}
In Eqs. ($\ref{efe}$) and ($\ref{ce}$), $a(t)$ is the scale factor
of the Universe, $H=\frac{\dot{a}(t)}{a(t)}$ is the Hubble parameter,
$\rho$ is the total energy density of the cosmic fluid, and $p$
is the pressure of the fluid. The ideal gas law is well known and
reads 
\begin{equation}
pV=nRT,\label{eos}
\end{equation}
where $p$, $V$, and $T$ are the pressure, volume and temperature
of the gas respectively. $n$ denotes the amount of substance in the
gas, often known as the number of moles, and $R$ is the universal
gas constant \footnote{In SI units it is $R=8.314~\text{\ensuremath{J/mol.K.}}$ }.
Now, if $m$ and $M$ denote the mass and the molar mass of the gas
respectively, and $\rho$ is the density, then we have 

\begin{equation}
V=\frac{m}{\rho}~~~~~~~~\text{and}~~~~~~~~n=\frac{m}{M},
\end{equation}
and the ideal gas law in Eq. ($\ref{eos}$) can be rewritten as 
\begin{equation}
p=R^{*}T\rho,\label{igeos}
\end{equation}
which we shall call the ideal gas equation of state. Here $R^{*}=\frac{R}{M}$
is known as the specific gas constant. Notice that the Universe undergoes
acceleration or deceleration according as $T<-\frac{1}{3R^{*}}$ or
$T>-\frac{1}{3R^{*}}$. This hints at a remarkable connection between
NAT and cosmic acceleration. NAT (or Kelvin temperatures) are interesting
and somewhat unusual, but not impossible or paradoxical {[}20 - 24{]}.
They are related to the concept of population inversion in statistical
physics {[}20 - 22{]}. The population inversion is obtained by a process
called optical pumping, which is a way of imparting energy to the
working substance of a laser in order to transfer the atoms to excited
states. Systems with a NAT will decrease in entropy as one adds energy
to the system \cite{key-23}. Most familiar systems cannot achieve
NATs because adding energy always increases their entropy. The possibility
of decreasing in entropy with increasing energy requires the system
to \textquotedbl{}saturate\textquotedbl{} in entropy \footnote{This implies that systems with NATs can never achieve thermodynamic
equilibrium.}, with the number of high energy states being small. These kinds of
systems, bounded by a maximum amount of energy, are generally forbidden
in classical physics. Thus, NAT is a strictly quantum phenomenon.
Hence, under special conditions (high-energy states are more occupied
than low-energy states), NAT are possible \cite{key-24}. These are
states existing in localized systems with finite, discrete spectra,
which can be prepared for motional degrees of freedom \cite{key-24}.
An intriguing example is the creation of an attractively interacting
ensemble of ultracold bosons at NAT that is stable against collapse
for arbitrary atom numbers \cite{key-24}. Another remarkable fact
is that NATs are hotter than all positive temperatures, even hotter
than infinite temperature (a proof is given in Appendix A of this
paper). 

Now, assuming no non-gravitational interaction, the temperature falls
\cite{key-25,key-26} with expansion as $1/a$ for relativistic fluids
such as photons (radiation) and as $1/a^{2}$ for non-relativistic
fluids like dust and cold dark matter. These expressions get only
slightly modified even if there is any interaction because such an
interaction is expected to be of very small strength. Thus, one can
write the expression for the temperature (in relativistic and non-relativistic
eras) in a compact form as 
\begin{equation}
T_{r,nr}=T_{\delta_{r},\delta_{nr}}a^{-(1+\delta)},\label{temp}
\end{equation}
where $\delta$ assumes the values $0$ and $1$ for relativistic
and non-relativistic eras respectively, and $T_{\delta_{r}}$ ($T_{\delta_{nr}}$)
represents the proportionality constant in relativistic (non-relativistic)
era.

Putting Eq. ($\ref{temp}$) into Eq. ($\ref{igeos}$) and using the
conservation equation ($\ref{ce}$), one gets the solution for $\rho$
(in relativistic and non-relativistic eras) as 
\begin{equation}
\rho_{r,nr}=\rho_{\delta_{r},\delta_{nr}}a^{-3}\text{exp}\left[\frac{3R^{*}T_{\delta_{r},\delta_{nr}}}{(1+\delta)}a^{-(1+\delta)}\right],
\end{equation}
where $\rho_{\delta_{r}}$ ($\rho_{\delta_{nr}}$) is the constant
of integration corresponding to relativistic (non-relativistic) era.
Note that $\rho_{r}\rightarrow\infty$ ($\rho_{nr}\rightarrow0$)
as $a\rightarrow0$ ($a\rightarrow\infty$). The deceleration parameter
for this model in relativistic and non-relativistic eras is given
by 
\begin{eqnarray}
q_{r,nr} & = & -\left(\frac{\dot{H}}{H^{2}}\right)_{r,nr}-1\nonumber \\
\\
 & = & \frac{1}{2}\left[1+3R^{*}T_{\delta_{r},\delta_{nr}}a^{-(1+\delta)}\right].
\end{eqnarray}
In terms of the redshift $z$, the above two expressions respectively
become 
\begin{eqnarray}
\rho_{r,nr}(z) & = & \rho_{\delta_{r},\delta_{nr}}(1+z)^{3}\text{exp}\left[\frac{3R^{*}T_{\delta_{r},\delta_{nr}}}{(1+\delta)}(1+z)^{(1+\delta)}\right],\\
q_{r,nr}(z) & = & \frac{1}{2}\left[1+3R^{*}T_{\delta_{r},\delta_{nr}}(1+z)^{(1+\delta)}\right].\label{qz}
\end{eqnarray}
Let us consider $z_{t}$ as being the redshift at which the Universe
transits from relativistic era to non-relativistic era. Then from
Eq. ($\ref{temp}$), we obtain $T_{r}=T_{\delta_{r}}(1+z_{t})$ and
$T_{nr}=T_{\delta_{nr}}(1+z_{t})^{2}$. Taking the ratio of these
two equations and noting that $T_{r}=T_{nr}$ at $z=z_{t}$, we get
\begin{equation}
z_{t}=\frac{T_{\delta_{r}}}{T_{\delta_{nr}}}-1.\label{zt}
\end{equation}
Then, the energy density and the deceleration parameter can be separately
expressed in the two eras as 
\begin{equation}
\rho(z)=\left\{ \begin{array}{lll}
\rho_{\delta_{r}^{'}}(1+z)^{3}\text{exp}\left[3R^{*}T_{\delta_{nr}}(1+z_{t})(1+z)\right]~~~(\text{relativistic era})\\
\\
\rho_{\delta_{nr}}(1+z)^{3}\text{exp}\left[\frac{3}{2}R^{*}T_{\delta_{nr}}(1+z)^{2}\right]~~~(\text{non-relativistic era}),
\end{array}\right.\label{rrnr}
\end{equation}
with 
\begin{equation}
\rho_{\delta_{r}^{'}}=\rho_{\delta_{nr}}\text{exp}\left[-\frac{3}{2}R^{*}T_{\delta_{nr}}(1+z_{t})^{2}\right],
\end{equation}
and 
\begin{equation}
q(z)=\left\{ \begin{array}{lll}
\frac{1}{2}\left\{ 1+3R^{*}T_{\delta_{nr}}(1+z_{t})(1+z)\right\} ~~~(\text{relativistic era})\\
\\
\frac{1}{2}\left\{ 1+3R^{*}T_{\delta_{nr}}(1+z)^{2}\right\} ~~~~~(\text{non-relativistic era}),
\end{array}\right.\label{qrnr}
\end{equation}
respectively.

Now, if $q_{t}$ ($\rho_{t}$) is the value assumed by the deceleration
parameter (energy density) at $z_{t}$, then, from either of the expressions
in Eq. ($\ref{qrnr}$), one gets 
\begin{equation}
R^{*}T_{\delta_{nr}}=\frac{2q_{t}-1}{3(1+z_{t})^{2}},
\end{equation}
and, by substituting in either of the expressions in Eq. ($\ref{rrnr}$),
one obtains 
\begin{equation}
\rho_{\delta_{nr}}=\rho_{t}(1+z_{t})^{-3}\text{exp}\left[-\frac{1}{2}(2q_{t}-1)\right].
\end{equation}
Thus, on one hand, one can determine the values of the unknown constants
$T_{\delta_{nr}}$ and $\rho_{\delta_{nr}}$ once $M$ is known. On
the other hand, $T_{\delta_{r}}$ can be obtained from Eq. ($\ref{zt}$).
Consequently, $\rho(z)$ and $q(z)$ can be exactly determined. Furthermore,
it is evident from the second expression in Eq. ($\ref{qrnr}$) that
$q(z)$ vanishes for two values of $z$:
\begin{equation}
viz.\;z=-1\pm\sqrt{\frac{1+z_{t}}{1-2q_{t}}},\label{eq: vision}
\end{equation}
provided $0<q_{t}<\frac{1}{2}$. In this regard, one also notes the
following points:
\begin{itemize}
\item $T_{\delta_{nr}}$ is negative which implies a NAT throughout the
non-relativistic era.
\item From Eq. ($\ref{zt}$), we observe that now $T_{\delta_{r}}$ should
assume a (negative) value less than $T_{\delta_{nr}}$ in order to
have a positive redshift.
\end{itemize}
The redshift $z=-1+\sqrt{\frac{1+z_{t}}{1-2q_{t}}}$ may correspond
to the transition of the Universe from the decelerating matter dominated
phase to the presently observed late time accelerating phase. However,
the other value of $z$ is smaller than $-1$ and, in turn, irrelevant
in the context of cosmic evolution. 

Thus, in a certain sense the NAT works like an ``effective quintessence''.
In fact, one can introduce an ``effective quintessential scalar field''
$\phi(z)$ starting from

\begin{equation}
\phi'(z)\equiv\frac{d\phi(z)}{dz}=\left\{ \begin{array}{lll}
-\frac{1}{(1+z)}\sqrt{1+\frac{2q_{t}-1}{3(1+z_{t})}(1+z)}~~~~~~~~~~~~~(\text{relativistic era})\\
\\
-\frac{1}{(1+z)}\sqrt{1+\frac{2q_{t}-1}{3(1+z_{t})^{2}}(1+z)^{2}}~~~~~(\text{non-relativistic era}),
\end{array}\right.\label{phidz}
\end{equation}
which implies

\begin{equation}
\phi(z)=\left\{ \begin{array}{lll}
\text{ln}\left(\frac{\sqrt{1+\frac{2q_{t}-1}{3(1+z_{t})}(1+z)}+1}{\sqrt{1+\frac{2q_{t}-1}{3(1+z_{t})}(1+z)}-1}\right)-2\sqrt{1+\frac{2q_{t}-1}{3(1+z_{t})}(1+z)}+C_{1}~~~~~~~~~~~~~(\text{relativistic era})\\
\\
\text{tanh}^{-1}\left(\frac{1}{\sqrt{1+\frac{2q_{t}-1}{3(1+z_{t})^{2}}(1+z)^{2}}}\right)-\sqrt{1+\frac{2q_{t}-1}{3(1+z_{t})^{2}}(1+z)^{2}}+C_{2}~~~~~(\text{non-relativistic era}),
\end{array}\right.\label{phiz}
\end{equation}
where $C_{1}$ and $C_{2}$ are arbitrary constants of integration.
The effective quintessential scalar field of Eq. (\ref{phiz}) is
associated to an ``effective potential''

\begin{equation}
\begin{array}{c}
V(\phi(z))=\\
\\
\begin{cases}
\begin{array}{lll}
\frac{1}{2}\rho_{t}(1+z_{t})^{-3}\text{exp}(1-2q_{t})(1+z)^{3}\text{exp}\left[\frac{2q_{t}-1}{1+z_{t}}(1+z)\right]\left\{ 1-\frac{2q_{t}-1}{3(1+z_{t})}(1+z)\right\} ~~~~~(\text{relativistic era})\\
\\
\frac{1}{2}\rho_{t}(1+z_{t})^{-3}\text{exp}\left(\frac{1}{2}-q_{t}\right)(1+z)^{3}\text{exp}\left[\frac{2q_{t}-1}{2(1+z_{t})^{2}}(1+z)^{2}\right]\left\{ 1-\frac{2q_{t}-1}{3(1+z_{t})^{2}}(1+z)^{2}\right\} ~(\text{non-relativistic era}).
\end{array}\end{cases}
\end{array}\label{vz}
\end{equation}
Then, one gets 
\begin{equation}
\rho(z)=\frac{1}{2}\dot{\phi}^{2}+V(\phi),
\end{equation}
with the associated Lagrangian 
\begin{equation}
\mathit{\mathfrak{L}}=\frac{1}{2}\partial_{\mu}\phi\partial^{2}\phi-V(\phi).
\end{equation}

\section{Some clarifications about the negative absolute temperature}

The negative temperature is the result of using the Boltzmann theorem
in order to calculate the system entropy \cite{key-32}. Only systems
whose their energy spectrum is bounded above can reach negative temperatures
\cite{key-32}. In this paper we use a classical gas in which energy
spectrum has not any restriction. Therefore, the attentive reader
could ask the following legitimate questions \cite{key-33}: What
is the origin of this negative temperature? Indeed, what is the maximum
bound of energy in this model? Can one consider the maximum value
of observable mass of universe in each cosmic era as the upper bound
for energy at that era? It is indeed possible to obtain a maximum
energy/mass bound in both relativistic and non-relativistic eras of
our model. Since $E=\frac{4}{3}\pi(ar)^{3}\rho$ and $T_{\delta_{r}}$,
$T_{\delta_{nr}}$ are both negative, one gets
\begin{equation}
E_{r}=\frac{4}{3}\pi(ar)^{3}\rho_{r}=4\pi a^{3}r^{3}\rho_{\delta_{r}}a^{-3}\mbox{exp}\left[3R^{*}T_{\delta_{r}}a^{-1}\right]<V\rho_{\delta_{r}}\label{eq: reply 1}
\end{equation}
for relativistic era, and
\begin{equation}
E_{nr}=\frac{4}{3}\pi(ar)^{3}\rho_{nr}=\frac{4}{3}\pi a^{3}r^{3}\rho_{\delta_{nr}}a^{-3}\mbox{exp}\left[\frac{3}{2}R^{*}T_{\delta_{nr}}a^{-2}\right]<V\rho_{\delta_{nr}}\label{eq: reply 2}
\end{equation}
for non-relativistic era. The upper bounds on $E_{r}$ and $E_{nr}$
can be exactly determined once the volume $V$ and the coefficients
$\rho_{\delta_{r}}$ and $\rho_{\delta_{nr}}$ are known.

It is also important clarify what negative absolute temperature means
in physical terms \cite{key-33}. Let us start with an analogy. In
order to boil water, it is required to add energy to it. During the
process of heating up, the water molecules gradually increase their
kinetic energy and move with faster average velocities. However, the
individual molecules possess different kinetic energies, ranging from
very slow to very fast. In other words, low energy states are more
probable in thermal equilibrium as compared to high energy states
which means that only few particles have very fast velocities. This
is referred to as Boltzmann distribution in physics. When particles
achieve negative absolute temperatures, then Boltzmann distribution
undergoes an inversion, i.e., most particles possess large energies,
while a few have small energies. As a consequence, a physical system
having negative temperature scale is hotter than any system with a
positive temperature. When a physical system having negative temperature
comes in touch with a physical system having positive temperature
heat flows from the negative temperature system to the positive temperature
system \cite{key-34,key-35}. This could appear as being a paradox,
but the problem is solved if one discusses temperature through the
thermodynamic rigorous definition of trade-off between energy and
entropy. In that case, the more fundamental quantity is the reciprocal
of the temperature, i.e. the thermodynamic beta. Hence, a system having
positive temperature increases in entropy if one adds energy to the
system. Instead, a system with having negative temperature decreases
in entropy when one adds energy to the system \cite{key-23}.

\section{Concluding Remarks}

In the approach of this work the cosmic fluid has been considered
as an ideal gas with a specific gas constant $R^{*}$ (to be fixed
by observations) but a varying NAT ($T_{r}$ in the relativistic era
and $T_{nr}$ in the non-relativistic era). Depending on whether the
temperature is greater (less) than $-\frac{1}{3R^{*}}$, the Universe
undergoes deceleration (acceleration). Hence, the cosmic evolution
from the offset of inflation to the present accelerating phase can
indeed be described by considering an ideal gas as the cosmic fluid.
In a certain sense, the NAT works like an ``effective quintessence''.
As a consequence, it is necessary to deal neither with exotic matter
like DE nor with modified gravity theories. Thus, the quantum nature
of the NAT might play an important role in studying the evolutionary
history of the Universe.

We observe that $q$ in Eq. ($\ref{qrnr}$) gives one transition at
a redshift $z_{p}=-\frac{1}{3R^{*}T_{\delta_{nr}}(1+z_{t})}-1$ in
the relativistic era which may correspond to the transition of the
early Universe from inflation to deceleration. In principle, one can
determine a lower bound on $T_{\delta_{nr}}$ by imposing the condition
$z_{p}>z_{t}$, which gives 
\begin{equation}
T_{\delta_{nr}}>-\frac{1}{3R^{*}(1+z_{t})^{2}}.
\end{equation}
Finally, if one again recalls the previously cited issue that NAT
states have been demonstrated in localized systems with finite, discrete
spectra, preparing a NAT state for motional degrees of freedom \cite{key-24},
one could shown that our atomic system is stable, even though the
atoms strongly attract each other, that means they want to collapse
but cannot due to being at NAT state. Thus, NAT implies negative pressures
and open up new parameter regimes for cold atoms, enabling fundamentally
new many-body states \cite{key-24}. The GTR shows that our Universe
as a whole is also not collapsing under the attractive force of gravity
\cite{key-19}, but it cannot explain that the Universe undergoes
an accelerated expansion. DE, which is believed to possess a huge
negative pressure, has been introduced to describe this effect. Therefore,
whether this negative pressure arises due to a negative fluid temperature
remains to be seen, and this is the same conclusion of \cite{key-24}.
It seems that NATs might play an important role in the dynamics of
our Universe and whether further investigations reveal a deep connection
between the nature of DE and the temperature of the cosmic fluid remains
to be seen.

For the sake of completeness, some clarifications on NATs have been
discussed in Section 3 of this work.

\section*{Acknowledgements}

Subhajit Saha was partially supported by SERB, Govt. of India under
National Post-doctoral Fellowship Scheme {[}File No. PDF/2015/000906{]}.
Anindita Mondal is thankful to DST, Govt. of India for providing Senior
Research Fellowship. Christian Corda is supported financially by the
Research Institute for Astronomy and Astrophysics of Maragha (RIAAM),
Project No. 1/4717-16.

The authors thank two unknown Reviewers for useful comments and advices.

\section*{Appendix A}

\textbf{Proposition:} \emph{Negative absolute temperatures are hotter
than positive absolute temperatures.}

\textbf{Proof:} Consider two bodies (1 \textbackslash{}\& 2) at different
temperatures ($T_{1}$ \textbackslash{}\& $T_{2}$) in contact with
one another. Suppose there is a transfer of a small amount of heat
$Q$ from body 1 to body 2, which changes the entropy of body 1 by
$-Q/T_{1}$ and that of body 2 by $Q/T_{2}$, so that the total change
in entropy is 
\[
dS=Q\left(\frac{1}{T_{2}}-\frac{1}{T_{1}}\right).
\]
 The above quantity must be positive according to the second law.
Now, if $T_{1}<0$ and $T_{2}>0$, then $\frac{1}{T2}-\frac{1}{T1}>0$,
which implies that body 1 (with a negative temperature) can transfer
heat to body 2 (with a positive temperature), but not the other way
around. Hence, negative absolute temperatures are hotter than positive
absolute temperatures.
\end{document}